\documentclass[apex]{jjap3}

\newcommand{\rmd}{\mathrm{d}}

\title{A Simple Method to Eliminate Shielding Currents Generated when Magnetization is Perpendicular to Superconducting Tapes Wound into Coils}
\author{Kazuhiro Kajikawa\thanks{E-mail address: kajikawa@sc.kyushu-u.ac.jp} and Kazuo Funaki}
\inst{Research Institute of Superconductor Science and Systems, Kyushu University, Fukuoka 819-0395, Japan}
\abst{Application of an external AC magnetic field parallel to superconducting tapes helps in eliminating the magnetization caused by the shielding current induced in the flat faces of the tapes.
This method helps in realizing a magnet system with high-temperature superconducting tapes for magnetic resonance imaging (MRI) and nuclear magnetic resonance (NMR) applications.
The effectiveness of the proposed method is validated by numerical calculations carried out using the finite-element method and experiments performed using a commercially available superconducting tape.
The field uniformity expected for practical applications is estimated to be less than 1 ppm.}

\begin{document}
\maketitle

Superconducting magnets used in magnetic resonance imaging (MRI) and nuclear magnetic resonance (NMR) require a highly uniform magnetic field at their center.
Hence, a multifilamentary superconducting wire with a circular or rectangular cross section composed of low-temperature superconductors such as NbTi or Nb$_3$Sn is wound around the magnets.
These materials are usually cooled by liquid helium, which has a boiling temperature of 4.2~K at atmospheric pressure.
However, because of the rapidly increasing demand for liquid helium and its limited availability, the cost of this coolant has been increasing steadily.
To realize a helium-free superconducting magnet system\cite{Glowacki}, it would be necessary to use liquid hydrogen or nitrogen, whose boiling temperatures are about 20~K and 77~K, respectively, as a coolant, or carry out conduction cooling with the help of a cryocooler.
Recently, a superconducting magnet with a nominal field of 0.5~T has been wound with a MgB$_2$ wire and installed in a private clinic for use in MRI\cite{Razeti}.
Nevertheless, designing a very high normal field (several tesla or several tens of tesla) magnet using the MgB$_2$ wire for MRI and NMR applications is very difficult because the critical current density decreases drastically when the magnetic field is on the order of a few tesla\cite{Tomsic}.
High-temperature superconductor (HTS) wires such as first-generation Bi-2223 Ag-sheathed tapes and second-generation Y-based or rare-earth-based coated conductors with a large length have been developed and made commercially available.
However, the superconducting parts of these wires have a cross section with a large aspect ratio: about 10 to 20 for the Bi-2223 tapes and several thousands to ten thousands for the coated conductors.
If a superconducting magnet wound with such a wire is energized, a given turn in the winding is exposed to a self-magnetic field, which arises from the transport current in the turn, and the external magnetic field generated by the currents in the other turns.
The combination of these magnetic fields causes inhomogeneity in the current flowing across the tape, and therefore, the magnetic field at the center of the magnet tends to be nonuniform\cite{Amemiya,Ahn,Miyazoe,Uglietti}.
Herein, we discuss the realization of a highly uniform magnetic field at the center of a superconducting magnet wound with tape-shaped HTS wires.

\begin{figure}[b]
\begin{center}
\includegraphics[scale=0.50]{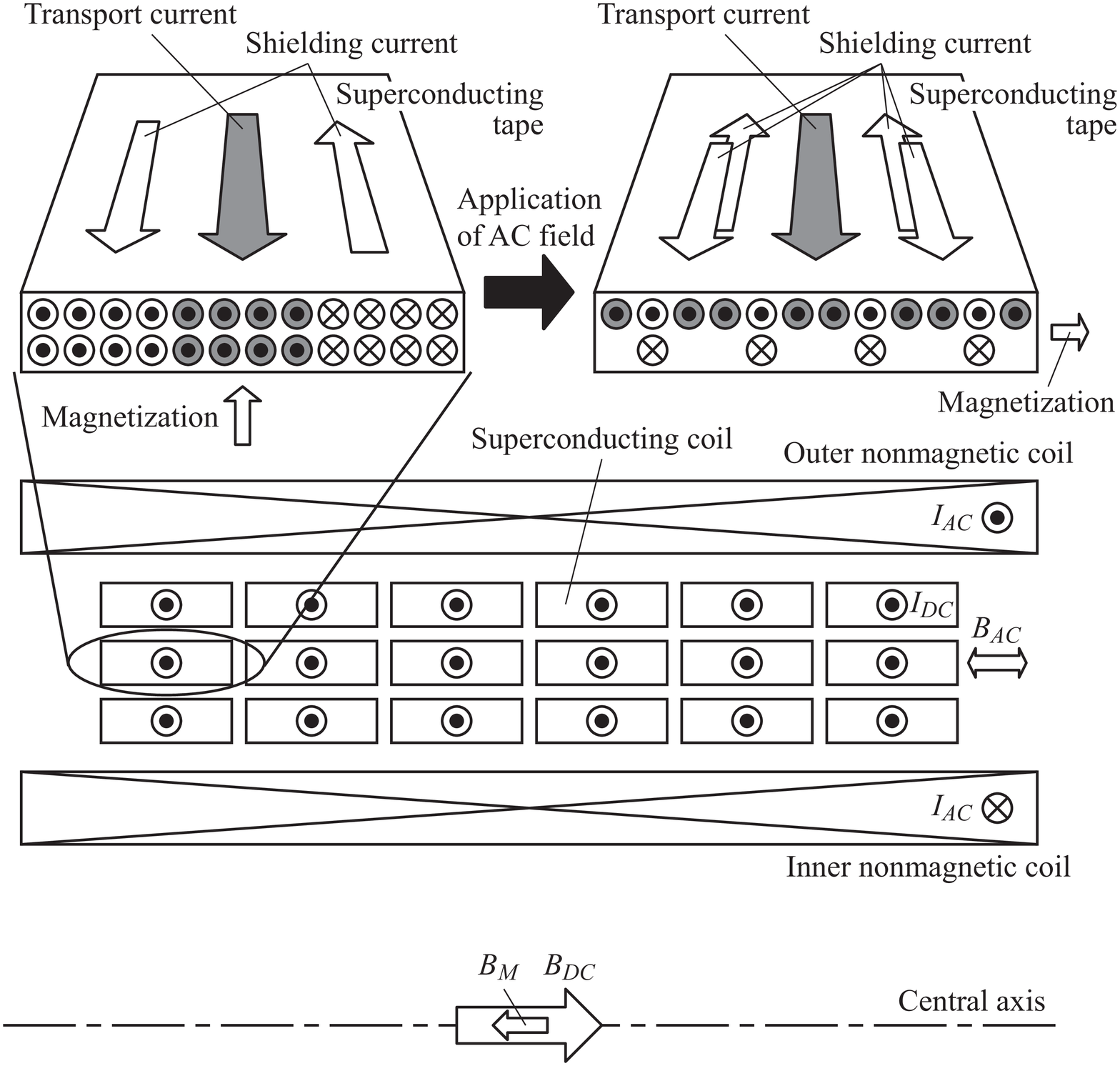}
\end{center}
\caption{Basic configuration of main coil wound with a superconducting tape and a pair of additional coils wound with nonmagnetic wires.
The magnetization perpendicular to the tape windings of the main superconducting coil can be eliminated by the AC magnetic field produced by the nonmagnetic coils.}
\label{fig1}
\end{figure}

Figure~\ref{fig1} shows the basic configuration of the magnet system proposed in this paper.
A superconducting coil is wound flatwise with HTS tapes, which are mainly magnetized perpendicular to their broad faces for the application of a direct transport current $I_{DC}$ to generate a magnetic field $B_{DC}$ at the center of the magnet.
The overall magnetization of all the superconducting tapes in the winding produces a magnetic field $B_M$ with very low uniformity inside the coil.
The superconducting coil is sandwiched between a pair of coils composed of nonmagnetic wires such as copper wires.
If these nonmagnetic coils are wound in opposite directions and energized by an alternating transport current $I_{AC}$ in a finite number of cycles, the HTS tapes constituting the windings are exposed to an external AC magnetic field $B_{AC}$, which is parallel to the flat faces of the tapes, for a short period.
If the amplitude of the external AC field is larger than the threshold value determined by the critical current density and thickness of the superconducting layer and the direct transport current in the tape, magnetization in the direction perpendicular to the AC field can be eliminated effectively.
This unique phenomenon of disappearance of the magnetization upon the application of the abovementioned type of external AC field, called ``abnormal transverse-field effect,'' was investigated systematically about thirty years ago for a linear array of monofilamentary superconducting wires\cite{Funaki_JJAP1,Funaki_JJAP2,Funaki_JJAP3,Funaki_KU}.
Recently, a similar phenomenon, ``vortex-shaking effect,'' was observed in the case of a superconductor strip\cite{Brandt_PRL,Brandt_SuST}.
The abnormal transverse-field effect (vortex-shaking effect) in the proposed configuration causes the current distribution in the width direction of the HTS tapes to become homogeneous.
Schematic images of the current profiles before and after the application of the external AC magnetic field parallel to the HTS tape are shown in the upper part of Fig.~\ref{fig1}; however, the local current cannot be clearly classified as the transport current or shielding current because it is usually given by the superposition of the transport and shielding currents.
Finally, the magnetization direction changes upon the application of the AC field, i.e., magnetization occurs in the direction parallel to the tape.
This parallel magnetization hardly affects the uniformity of the central magnetic field of the coil, as will be shown later in the document.

\begin{figure}[b]
\begin{center}
\includegraphics[scale=1.00]{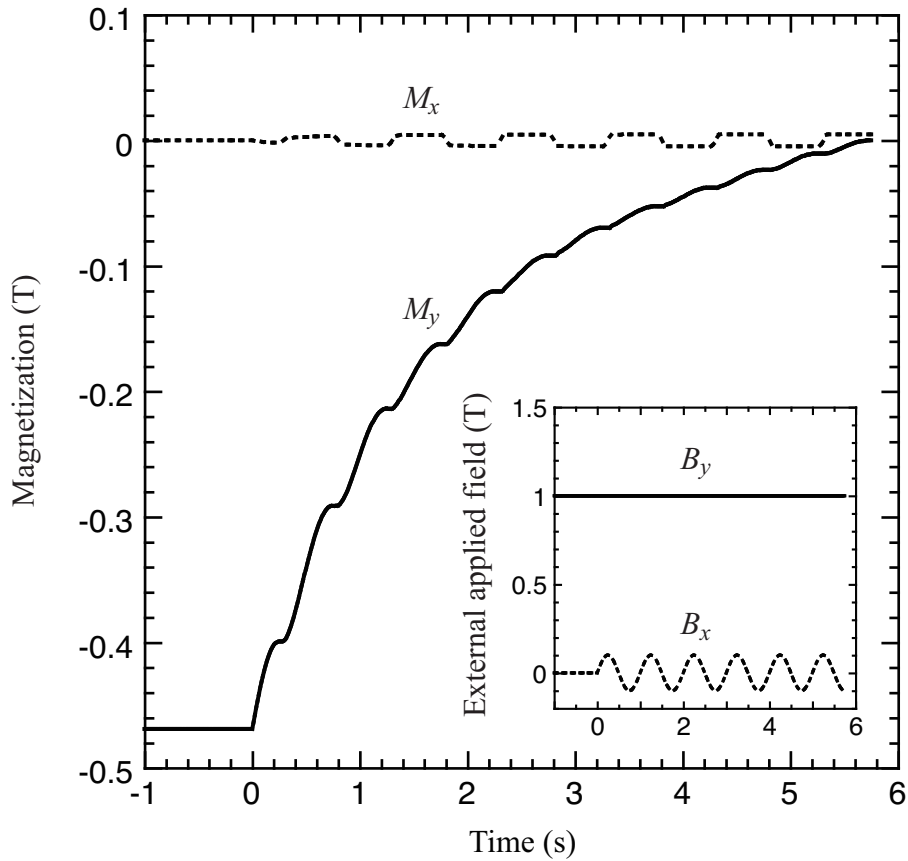}
\end{center}
\caption{Time evolution of magnetization in the superconducting tape calculated numerically by the finite-element method.
The inset shows the waveforms of the external applied magnetic fields.}
\label{fig2}
\end{figure}

In order to evaluate the decrease in the magnetization perpendicular to the superconducting tape and the homogeneous distribution of the transport current in the width direction on the basis of the abnormal transverse-field effect, numerical calculations are carried out using the two-dimensional finite-element method formulated with the self-magnetic field resulting from the currents induced in the analysis region, which is discretized into edge elements\cite{Kajikawa_IEEE-TAS,Kajikawa_JSNM}.
The width $2a$ and thickness $2b$ in the $x$- and $y$-directions of the superconducting tape with an infinite length in the $z$-direction are assumed to be 10~mm and 0.1~mm, respectively, and therefore, the aspect ratio of the cross section is 100.
This assumption helps in carrying out efficient numerical analysis even with a limited number of computer resources and validating the proposed method.
Further, the relationship between the electric field and the local current density in the superconducting tape is represented by Bean's critical-state model\cite{Bean,London}, in which the critical current density is assumed to be independent of the local magnetic field.
The critical current density is fixed at 2~$\times$~10$^8$~A/m$^2$, and therefore, the critical current is 200~A.
The flux-flow state over the critical current density is taken into account along with the flux-flow resistivity of 10$^{-7}$~$\Omega\cdot$m\cite{Kajikawa_IEEE-TAS}.
An example of the numerical results is shown in Fig.~\ref{fig2}, where the superconducting tape has a direct transport current of 100~A.
The vertical axis in this figure represents the magnetization of the superconducting tape in the $x$- and $y$-directions, $M_x$ and $M_y$, respectively, which are estimated from the following expressions:
\begin{equation}
M_x=\dfrac{\mu_0}{4ab}\int_{-b}^{b}\rmd y\int_{-a}^{a}yJ\!\left(x,y\right)\rmd x,\label{eq1}
\end{equation}
\begin{equation}
M_y=-\dfrac{\mu_0}{4ab}\int_{-b}^{b}\rmd y\int_{-a}^{a}xJ\!\left(x,y\right)\rmd x.\label{eq2}
\end{equation}
Although it is difficult to evaluate the magnetization using the aforementioned equations for a long superconductor with a finite transport current, the equations may be credible if the origin of the coordinate system is set at the center of the cross section of the tape.
Under this assumption, these equations become zero for the tape carrying the critical current, and they can be used exactly for no transport current.
As shown in the inset of Fig.~\ref{fig2}, an external DC magnetic field $B_y$ of 1~T in the $y$-direction is first applied to the superconducting tape, and subsequently, the tape is exposed to an external AC field $B_x$ with an amplitude of 0.1~T and a frequency of 1~Hz in the $x$-direction.
The superconducting tape is first magnetized significantly in the DC field, and this magnetization decreases exponentially upon the application of the AC field.
Moreover, magnetization occurs in the direction of the AC field despite the existence of the DC field.
The numerical calculation was stopped when the absolute value of the magnetization became lower than 10$^{-4}$~T in the $y$-direction, implying that in the proposed method, the perpendicular magnetization was reduced by at least three orders of magnitude from the initial value.
According to the final result obtained just before stopping the calculation, the standard deviation of the sheet current distribution along the tape width was 0.1\%.

\begin{figure}[b]
\begin{center}
\includegraphics[scale=1.00]{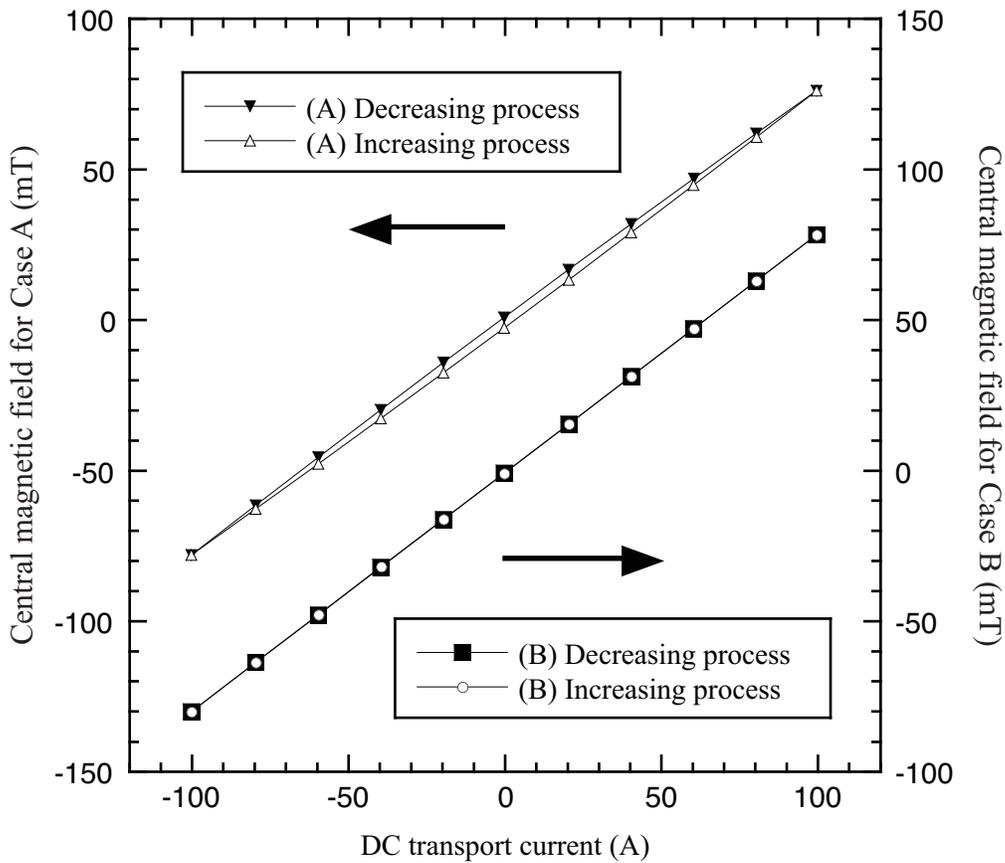}
\end{center}
\caption{Experimental results obtained for the central magnetic field in the superconducting coil with fixed DC transport currents.
Cases A and B represent the central fields without and with the application of an external AC magnetic field at every step, respectively.
The solid lines are guides to the eye.}
\label{fig3}
\end{figure}

A set of small coils is fabricated to validate the method used for nullifying the influence of magnetization on the central magnetic field of a coil wound with the HTS tape.
A superconducting solenoid coil is wound with a commercially available GdBa$_2$Cu$_3$O$_x$ tape with a width of 5~mm and a superconducting layer with a thickness of 1.5~$\mu$m.
The critical current of this tape is 224~A at 77~K and self-field.
The number of layers, turn number, inner diameter, outer diameter, and mean height of the coil are 4, 70.5, 63.0~mm, 72.1~mm, and 89.5~mm, respectively.
A couple of nonmagnetic coils are also prepared using a copper wire of 1~mm diameter.
The turn numbers and heights of the prepared coils are identical, 264 and 139.8~mm, respectively.
The inner and outer diameters of the inner copper coil are 53.0~mm and 57.0~mm, respectively.
On the other hand, the inner and outer diameters of the outer coil are 79.0~mm and 82.9~mm, respectively.
The HTS coil and copper coils are located coaxially and immersed in liquid nitrogen.
Figure~\ref{fig3} shows a set of examples of the experimental results.
Only the HTS coil is energized in Case A.
A bipolar power source is connected to the HTS coil to supply current, which is first monotonically decreased from 100~A to $-100$~A and subsequently increased up to 100~A in steps of 20~A.
The Hall probe placed at the center of the coils is used to observe the magnetic field in the axial direction for each value of direct transport current applied to the HTS coil.
In Case A, the central fields for the applied currents are irreversible and hysteretic owing to the shielding currents induced in the HTS tapes.
In Case B, the fabricated copper coils are used in addition to the HTS coil; the two copper coils wound in opposite directions are connected in series to an AC power supply and excited by an alternating transport current with an amplitude of 30~A and a frequency of 100~Hz for 30~s after every current-holding step for the HTS coil.
In this case, the hysteretic behavior is not observed and the central magnetic fields become more linear.
The deviations of the central magnetic fields from an approximate line of the experimental results are estimated to be on the order of several percents for Case A, but these deviations are less than 0.1\% for Case B.
The value in the latter case is very close to the lower limit of measurement in the present experimental setup.

The influence of magnetization parallel to the superconducting tapes in a single-layer solenoid coil after the application of an external AC magnetic field on the field profile inside this coil is evaluated.
The superconducting coil is assumed to be a very thin cylinder with a mean radius and height of $R$ and $2h$, respectively.
If the coil is closely wound, the thickness of the cylinder will be given by the thickness of superconducting layer, $2b$.
Since magnet systems used for MRI and NMR are usually composed of a main coil and several shim coils to ensure a highly uniform magnetic field at the center of the magnet, their design is complicated.
For estimating the influence of parallel magnetization on the inside magnetic field, the deviation of the field profile along the central axis from the nonmagnetization case is considered\cite{Kajikawa_ISS97}:
\begin{equation}
\dfrac{B_M}{B_{DC}}\simeq\dfrac{b\!\left(1-i^2\right)}{2R\!\left(\beta^2+1\right)\!i}\!\left[1+\dfrac{3\!\left(\beta^2-1\right)}{\left(\beta^2+1\right)^2}\zeta^2\right]\;\mathrm{for}\;\zeta\ll 1,\label{eq3}
\end{equation}
where $i=I_{DC}/I_c$, $\beta=h/R$, and $\zeta=z/R$.
The origin of the axial coordinate $z$ is set at the center of the coil.
In a typical case where $b=1$~$\mu$m, $R=0.5$~m, $\beta=2$, and $i=0.5$, Eq.~(\ref{eq3}) becomes $B_M/B_{DC}\simeq3\times10^{-7}\left(1+0.36\,\zeta^2\right)$.
This implies that the proposed method is very promising for realizing a magnet system composed of HTS tapes, with a highly uniform central magnetic field, for use in MRI and NMR.
To obtain a more uniform field, the multilayer windings are separated into two coaxial parts, and external AC fields are applied to these parts in opposite directions.
In this case, the consequent bidirectional magnetizations parallel to the tapes cancel out the small derivation of the central magnetic field.
The proposed method to eliminate the perpendicular magnetization of HTS tapes would also be applicable to other magnet systems requiring a highly uniform inside magnetic field such as a particle accelerator.

\end{document}